# Surfactant-Free Synthesis of Spiky Hollow Ag-Au Nanostars with Chemically Exposed Surfaces for Enhanced Catalysis and Single-Particle SERS

Ziwei Ye[†a], Chunchun Li[†a], Maurizio Celentano[a], Matthew Lindley[b], Tamsin O'Reilly[a], Adam J. Greer[c], Yiming Huang[a], Christopher Hardacre[c], Sarah J. Haigh[b], Yikai Xu*[a] and Steven E. J. Bell*[a]

[a] Dr. Z. Ye, Dr. C. Li, Dr. M Celentano, T. O'Reilly, Y. Huang, Dr. Y. Xu, Prof. S. E. J. Bell

School of Chemistry and Chemical Engineering, Queen's University of Belfast, University Road, Belfast, BT9 5AG, Northern Ireland (UK)

[b] M. Lindley, Prof. S. Haigh

Department of Materials, The University of Manchester, Oxford Road, Manchester, M13 9PL

[c] Dr. A. J. Greer, Prof. C. Hardacre,

Department of Chemical Engineering & Analytical Science, The University of Manchester, Oxford Road, Manchester, M13 9PL

*KEYWORDS surfactant-free, spiky nanoparticles, hollow nanoparticles, SERS, cubic core*

**ABSTRACT:** Spiky/hollow metal nanoparticles have applications across a broad range of fields. However, current bottom-up methods to produce spiky/hollow metal nanoparticles rely heavily on the use of strongly adsorbing surfactant molecules, which is undesirable since these passivate the product particles' surfaces. Here we report a high-yield surfactant-free synthesis of spiky hollow Au-Ag nanostars (SHAANs). Each SHAAN is composed of >50 spikes attached to a hollow ca. 150 nm diameter cubic core, which makes SHAANs highly plasmonically and catalytically active. Moreover, the surfaces of SHAANs are chemically exposed which gives them significantly enhanced functionality compared to their surfactant-capped counterparts, as demonstrated in surface-enhanced Raman spectroscopy (SERS) and catalysis. The chemical accessibility of the pristine SHAANs also allows the use of hydroxyethyl cellulose as a weakly-bound stabilizing agent. This produces colloidal SHAANs which remain stable for >1 month while retaining the functionalities of the pristine particles and allow even single-particle SERS to be realized.

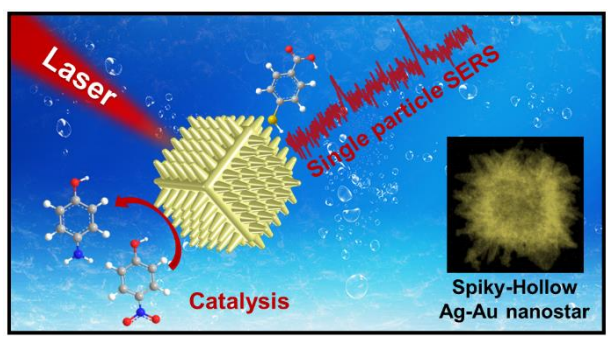

## Introduction

Ag and Au nanoparticles have been extensively studied in the past few decades due to their unique catalytic and plasmonic properties.[1-6] Up to now, the most popular route to produce Ag and Au nanomaterials is through bottom-up colloidal synthesis, which typically employs a strongly adsorbing surfactant to direct particle growth and/or provide colloidal stability.[7-11] A variety of anisotropic metal nanoparticles with distinct properties can now be synthesized with the aid of surfactants, for instance, polyvinylpyrrolidone (PVP)-assisted synthesis of nanocubes;[12-14] cetyltrimethylammonium bromide (CTAB)-assisted synthesis of nanorods;[15-17] Triton-X-assisted synthesis of nanostars etc.[18-20] While the use of surfactant capping molecules offers easy manipulation of the morphology of synthesized nanoparticles, their existence are also problematic for nearly every type of application, for example in plasmonic sensing and catalysis, where the surface-bound surfactant molecules act as both physical and chemical barriers that restrict the free access of analytes/reactants to the surface of nanoparticles,[9,21-23] or in bio-applications where the free surfactants induce cytotoxicity.[24-25] As a result, methods for post-synthesis removal of surfactant molecules or surfactant-free synthesis have become important areas of research.[26-30] For example, Gao et al showed that diethylamine can be used as a general and highly effective intermediate ligand to facilitate the replacement of strongly adsorbed capping agents with weakly adsorbed capping agents.[29] Odom et al demonstrated that it was possible to use weakly adsorbing and biocompatible Good's buffers, such as 4-(2-hydroxyethyl)-1-piperazineethanesulfonic acid, to synthesize Au nanostars.[30]

Among the anisotropic metal nanoparticles reported to date, spiky or hollow nanoparticles have drawn particular interest as catalysts due to their large surface-to-volume ratio and the presence of high-index crystal planes which are composed of highly active low-coordinated atoms.[31-32] If they are composed of plasmonic materials, such as Au/Ag, spiky or hollow nanoparticles have also been shown to possess significantly enhanced plasmonic properties compared to their isotropic counterparts.[33-36] To take full advantage of the plasmonic and catalytic properties of hollow and spiky nanoparticles a few research groups have reported methods to synthesize Au/Ag nanostars which are both spiky and hollow.[20,37-41] However, in general, bottom-up methods which allow the production of spiky Au/Ag nanostars that contain a hollow interior remain extremely rare. More importantly, current methods for producing Ag/Au nanostars with hollow interiors are either complex or require extensive use of surfactants, which have significantly limited the application of these particles, despite their promising properties. For example, Rodríguez-Fernández et al showed that spiky and hollow Au nanoparticles with tailored near-infrared plasmonic properties could be produced using Triton X-100 as the growth directing agent.[20] Evans et al showed that spiky and hollow Au nanoparticles with excellent catalytic and SERS activity could be produced using methyl-orange-$FeCl_3$ templates.[38]

In this work, we demonstrate a mild and surfactant-free synthesis, which can be completed at room-temperature within minutes to produce spiky hollow Au-Ag nanoparticles (SHAANs) with ca. 100% morphological yield. Importantly, the particle surfaces of the product SHAANs are chemically exposed which meant that they could be stabilized using extremely weakly-bound capping ligands, such as hydroxyethyl cellulose (HEC); this allowed the product colloid to remain stable for >1 month while retaining the surface-accessibility of the particles. The unique morphology of the SHAANs combined with their highly accessible surfaces allow them to exhibit outstanding catalytic and single-particle SERS activity which were significantly enhanced compared to their conventional surfactant-capped counterparts. The ability to access enhanced nano-properties with anisotropic nanoparticles prepared using an easy and reliable room temperature process not only offers a powerful tool for fundamental research but also paves the way for a broad range of important real-life applications, such as trace molecule detection, biomedical diagnostics or the construction of functional devices.

**Results and Discussion**

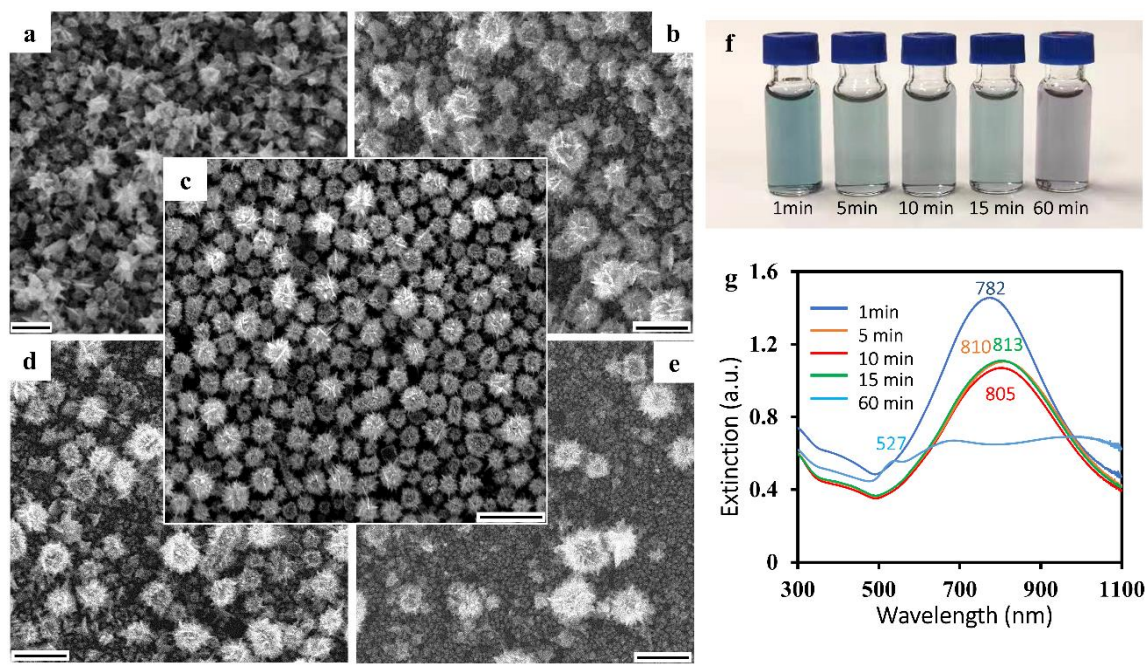

**Figure 1.** (a)-(e) Scanning electron microscopy (SEM) images of Au/Ag alloy nanoparticles synthesized by incubating the mixture solution of $HAuCl_4$, NaCl and $AgNO_3$ for 0 min (a), 5 min (b), 10 min (c), 15 min (d), 60 min (e) and then adding ascorbic acid. Scale bars in all images correspond to 500 nm. (f) Corresponding optical images of colloids prepared with different incubation times. (g) UV-vis spectra of colloids prepared with different incubation time.

The synthesis of the SHAANs is a simple and rapid one-step process. Briefly, NaCl and $AgNO_3$ solution are first added in sequence to a $HAuCl_4$ solution under constant stirring. The mixture is then incubated after which L-ascorbic acid solution is quickly injected (see details in the Experimental section of SI). This causes the color of the solution to change immediately (typically within ca. 15s), indicating the formation of the product nanoparticles. The critical step in the process is the incubation of mixed $HAuCl_4$, NaCl and $AgNO_3$ solution before the addition of L-ascorbic acid. As shown in **Figure 1a-e**, the morphology of the nanoparticles obtained changed dramatically with different incubation times. The addition of ascorbic acid without incubation produced small spiky nanoparticles with an average diameter of ca. 40 nm. The resulting colloid had a dark blue color with $\lambda_{max}$ measured via UV-vis at ca. 782 nm (Figure 1f-g). Further scanning transmission electron microscopy (STEM) characterization and energy dispersive X-ray (EDX) spectroscopy of this colloid sample revealed that some of the nanoparticles which appeared to have a hollow interior when observed with scanning electron microscopy (SEM) actually had a Ag/AgCl core (Figure S1). As shown in Figure 1b, when the incubation time was extended to 5 min, approximately half of the nanoparticles became larger SHAANs with a particle diameter of c.a. 150 nm. These SHAANs had a larger number of spikes on the surface compared to smaller nanoparticles obtained with less incubation and the spikes were notably narrower and longer. Additionally, the interiors of particles were found to be hollow, which will be discussed in detail in the following section. The change in morphology with incubation was also reflected in the color of the colloids and their UV-vis extinction spectra. As shown in Figure 1f-g, the color of the colloids with incubation was clearly different and their extinction peaks were broadened and red-shifted significantly compared to the non-incubated sample. The optimal incubation time, which led to the highest yield (ca. 100% morphological yield) of SHAANs, was found to be 10 min, as shown in Figure 1c. The resulting colloid was grey with its UV-vis extinction peak centred around 805 nm (Figure 1f-g). Extending the incubation time beyond 10 min led to the population of SHAANs decreasing and eventually nearly disappearing completely. As shown in Figure 1d, the population of the product colloid with 15 min of incubation was similar to those obtained with 5 min of incubation, which is reflected by their identical color and UV-vis extinction (Figure 1f-g). If the incubation time was increased up to 60 min, very few SHAANs were formed and the product colloid consisted mainly of small (ca. 30 nm) spheres (Figure 1e). Unsurprisingly, this colloid was purple with a UV-vis extinction centered at around 527 nm (Figure 1f-g), characteristic of small spherical Au nanoparticles.[42]

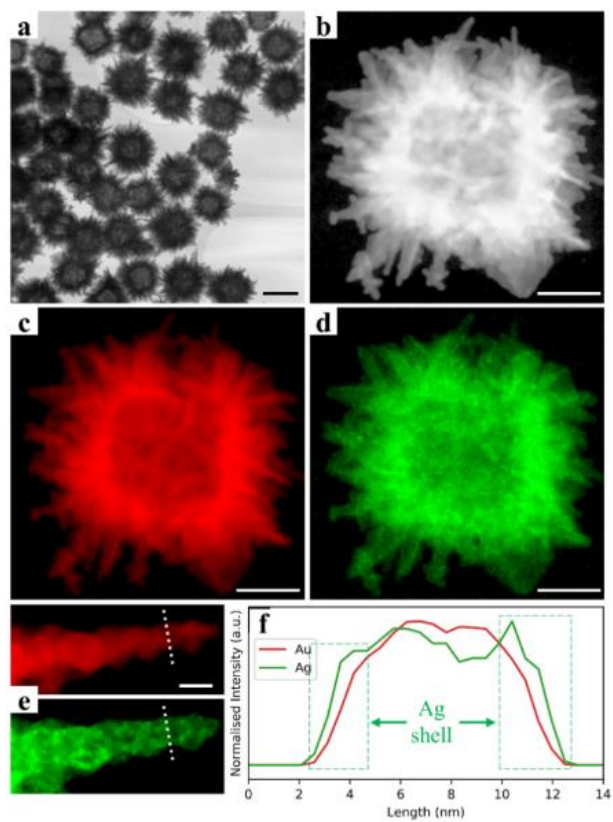

**Figure 2**. (a) Low and (b) high magnification scanning transmission electron microscopy high-angle annular dark field (STEM-HAADF) images of SHAANs. Energy dispersive X-ray (EDX) elemental mapping of SHAANs: (c) Au mapping image; (d) Ag mapping image. (e) EDX elemental mapping of a tip on the SHAAN shown in (b). (f) Plot showing the materials composition obtained with high-magnification elemental line-scan across the line labelled in (e). Scale bar in (a), (b-d) and (e) correspond to 200 nm, 60 nm and 10 nm, respectively.

Further characterization of the nanoparticles was performed by scanning transmission electron microscopy high-angle annular dark field (STEM-HAADF) imaging. This was used in combination with EDX spectroscopy to investigate the internal morphology, local elemental distribution and composition of the SHAANs. Images from the 10-minute incubated specimen revealed that the homogeneous population of nanoparticles have cubic hollow cores in addition to the spiky surface morphology observed by SEM (**Figure 2a**). The average diameter of the cubic core was found to be 150±16 nm as measured from 100 individual SHAANs. Figure 2b shows the high magnification STEM-HAADF image of a typical SHAAN. The hollow cavity of the SHAAN was measured to be ca. 100 nm in diameter, and the length of spikes ranged between 30-45 nm. Elemental mapping revealed a fairly evenly dispersed mixture of Ag and Au, which resembled bimetallic Au-Ag alloys, with a mean composition of 90 $_{at.}$% Au and 10 $_{at.}$% Ag (Figure 2c-d). Interestingly, as shown in Figure 2 e-f, the high-magnification elemental line-scan also revealed that the $_{at.}$% ratio between Ag and Au was clearly higher at the surface of the SHAANs, which suggested the existence of a ~1nm thick Ag-rich surface layer that could have been formed due to the lower surface energy of Ag compared to Au.[43] STEM-HAADF analysis for a non-incubated specimen (see Figure S2) also revealed some hollow particles but with a smaller and less regular size and shorter spikes, consistent with the SEM data. These particles also had a composition of 90 $_{at.}$% Au and 10 $_{at.}$% Ag and a few nanometer thick Ag-rich surface layer (Figure S3-4).

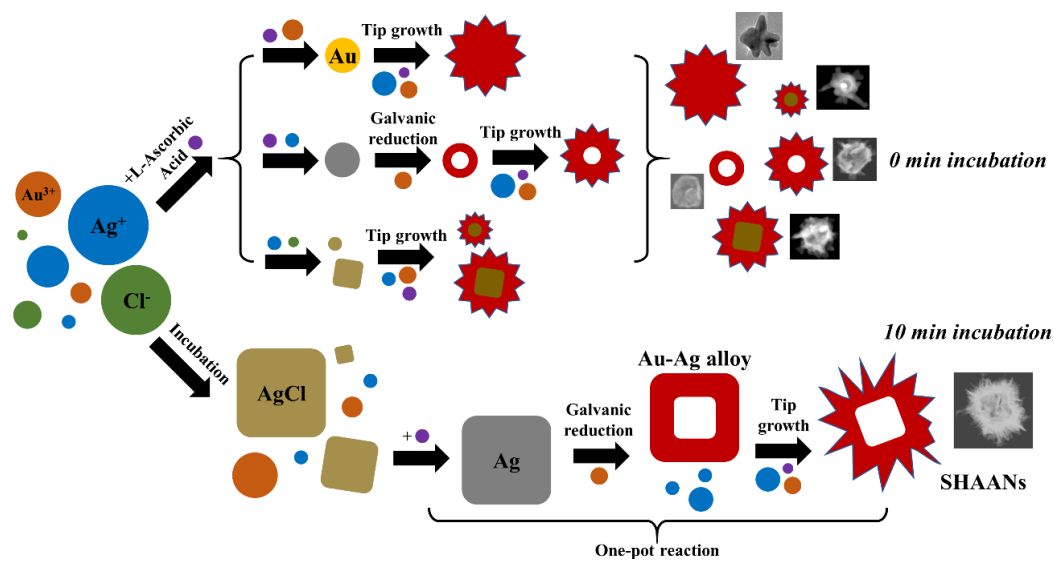

**Scheme 1.** Schematic illustration of the formation of SHAANs (10 min incubation time) compared to with 0 min incubation time.

 The proposed mechanism for the formation of SHAANs is shown in **Scheme 1**. The synthesis is initiated by addition of AgNO$_3$ to a mixed NaCl/HAuCl$_4$ solution, which leads to the formation of small AgCl seed nanoparticles. This is consistent with the observation that the mixture solution changes gradually from transparent to slightly cloudy after the addition of AgNO$_3$. Subsequent incubation leads to the growth of the AgCl seeds to form larger cubic AgCl nanoparticles. However, this process can be interrupted by addition of ascorbic acid, which triggers a series of redox reactions (Table S1) that ultimately led to formation of SHAANs if the optimized incubation time of 10 mins is used. More specifically, the addition of ascorbic acid to the incubated solution leads to the reduction of the AgCl nanoparticles to Ag$^0$ nanoparticles, as has been previously observed.[44-46] However, these are not stable in the presence of Au$^{3+}$ cations, which can galvanically reduce them, leading to the replacement of the Ag$^0$ by Au$^0$ in the nanocrystal structure, a process which would be expected to yield cubic hollow Au shells.[44-45] Simultaneously, the presence of ascorbic acid and Au$^{3+}$ cations along with residual Ag$^+$ and Cl$^-$ leads to growth of spikes on the surface of the particles.[39]

This mechanism accounts for the importance of the duration of the incubation step and the nature of the products which are formed when it is changed from the optimum value. If the initial mixture was over-incubated (60 min), the majority of Ag$^+$ and Cl$^-$ ions were consumed in the growth of AgCl nanoparticles and not enough were left to assist the formation of Au spikes. Therefore, the nanoparticles formed by subsequent addition of ascorbic acid were mostly quasi-spherical (Figure 1e). In contrast, if the initial mixture was under-incubated (0 min), there are not sufficient AgCl seeds to template the growth of SHAANs and the AgCl seeds that are available are under-grown, which results in the formation of poly-disperse spiky/hollow nanoparticles (Figure 1a and Figure S2). Although it is important to use the correct incubation time to achieve appropriate balance between AgCl seeds, Ag$^+$ and Cl$^-$ necessary for the formation of SHAANs (Figure 1c), it is useful to stress that provided this time is used the process is very robust and it reliably generates SHAANs, as shown by SEM and SERS measurements in Figure S5-6. Although AgCl nanoparticles act as intermediates during the synthesis, a significant amount of residual AgCl is not expected to be present in the product SHAANs because it is reduced by the excess ascorbic acid in the solution. This was confirmed by the observation that washing SHAANs with ammonia water, which would dissolve any residual AgCl, gave no detectable change in the morphology of the SHAANs (Figure S7).

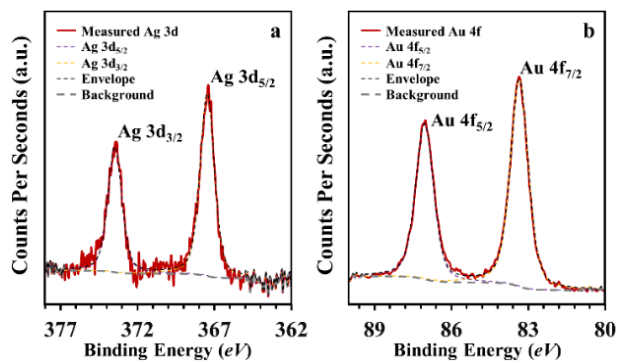

**Figure 3**. XPS analysis of SHAANs showing the characteristic peaks of (a) Ag and (b) Au.

X-ray photoelectron spectroscopy (XPS) was used to further study the surface composition of the SHAANs, in particular to identify the oxidation state of the metals and the number of species present. Surprisingly, the XPS revealed that Ag was present in just one oxidation state, $Ag^+$, indicated by the photoelectron peak at 367.5 eV (Ag $3d_{5/2}$) (Figure 3a).[47-48] Since, as discussed above, it is unlikely that a significant amount of AgCl is present in the product SHAANs, the $Ag^+$ observed by XPS is most likely in the form of $Ag_2O$, which is formed post-synthesis due to the high reactivity of colloidal Ag towards oxygen.[49] As shown in Figure 3b, the XPS also showed a photoelectron peak at 83.4 eV, corresponding to metallic Au $4f_{7/2}$. Interestingly, the binding energy observed for Au was lower than conventional values reported in literature (84.0 eV). This could be due to charge-transfer effects which arise from the formation of Ag-Au alloys in the SHAANs.[50-51] Table S2 and Figure S8 shows the XPS data and at.% of the relevant elements in the SHAANs sample from XPS. Importantly, as a surface-specific technique the XPS results were consistent with the Ag surface enrichment observed in the STEM-elemental mapping, and showed that the at.% ratio between Ag and Au (1:1.6) on the surface of the SHAANs was much higher than the overall ratio between Ag and Au in the whole particle (1:9).

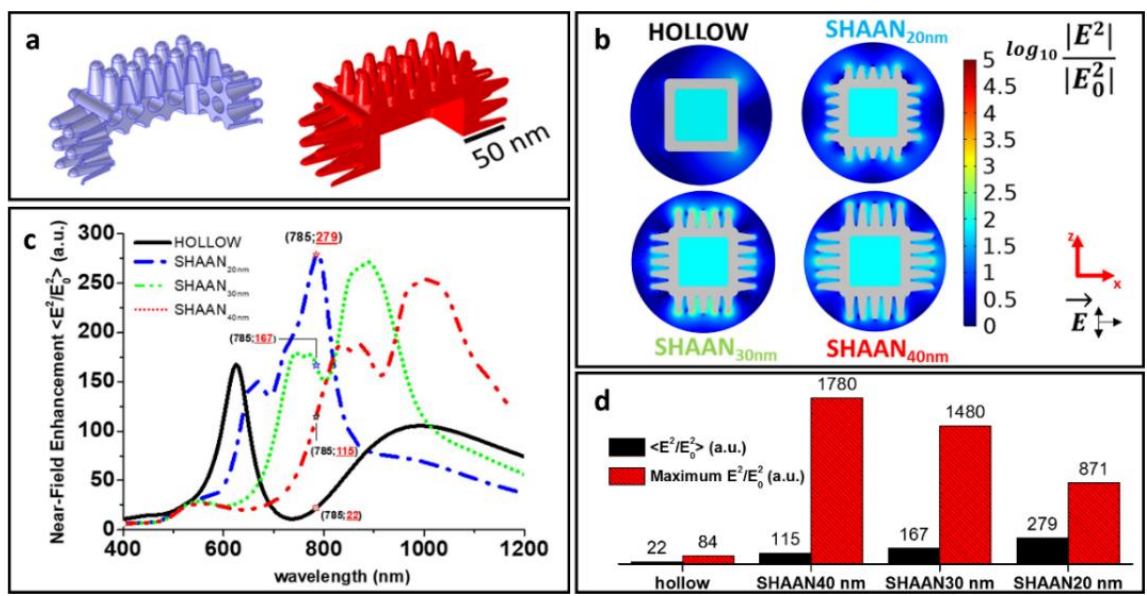

**Figure 4**. Finite element method simulation of the electrodynamic of SHAANs$_x$ and hollow cube nanostructures. (a) 3D renderings of a SHAAN, in red (30 nm spike length), and of the volume, in light blue, which defines the volume average near-field enhancement factor; (b) near-field enhancement factor maps for SHAANs$_x$ with three different spike length $x$ ($x$ = 20, 30 or 40 nm) and the hollow nanocube core ; (c) volume average near-field enhancement factors for the structures in (b); (d) comparison between the volume average near-field enhancement factors and maximum punctual enhancement factor of the structures in (b) calculated at 785 nm excitation wavelength.

The plasmonic properties of SHAANs were investigated using a finite element method, with the particle structure and material composition modelled according to the microscopy measurements (see Experimental in SI). The volume average near-field enhancement factor was calculated as:

$$\left< \frac{|E^2|}{|E_0^2|} \right> = \left( \iiint \frac{|E^2|}{|E_0^2|} dV \right) / V$$

where the volume V was made by a 2 nm thick shell covering the nanoparticle surface (**Figure 4a**). Figure 4b compares the calculated electromagnetic field distribution of individual SHAANs$_x$ ($x$ corresponding to the approximated length of the spikes within an SHAAN) and a hollow cube core excited by 785 nm laser. Regardless of the exact spike length, the near-field enhancement factor maps showed that the distribution of the excited surface plasmon resonance was mainly localized around the sharp tips, rather than within the hollow cavity. Essentially, the electromagnetic-field distribution around the SHAANs was much more similar to that of solid nanostars rather than hollow/porous nanoparticles or nanocubes.[52-54] The profile and strength of the volume-averaged near-field enhancement was found to be highly dependent on the length of the spikes. For example, Figure 4c shows a plot comparing the volume-averaged near-field enhancement of individual hollow cubes and SHAANs$_x$ at different wavelengths, which showed that the optimal excitation wavelength shifted by >200 nm when the spike length increased from 20 to 40 nm. Figure 4d compares the maximum and average near-field enhancement of hollow nanocubes and SHAANs$_x$ at the same excitation wavelength as our Raman measurements (see below). As shown in the plot, the SHAANs$_x$ were significantly more plasmonically active compared to hollow nanocubes. Interestingly, the simulations also revealed that the maximum near field enhancement increased with increasing spike-length of the SHAANs while the volume-averaged enhancement decreased.

The SERS performance of SHAANs was first investigated using 4-mercaptobenzoic acid (MBA) as the model analyte. Strong SERS signals of MBA were observed using freshly prepared, well-dispersed pristine SHAANs colloids as the enhancing substrate and the limit of detection was measured to be ca. $10^{-8}$ M, as shown in Figure S9. The "Analytical Enhancement Factor" (AEF), which is calculated differently to the Enhancement Factor (EF), is widely regarded as the gold standard for quantifying the enhancing ability of colloidal SERS substrates since the complex geometry makes directly calculating EF difficult.[55-56] Using MBA as the probe analyte the AEF of SHAANs was determined to be ca. $4.9 \times 10^5$ under non-resonant conditions, for the pristine SHAANs colloid (Figure S9). This value is notably stronger than the enhancement predicted in Figure 4, which is likely due to the presence of additional far-field electromagnetic enhancement and chemical enhancement that is not considered in the simulations. It is worth noting that this experimentally recorded value is comparable to the AEF value obtained from aggregated isotropic nanoparticles, such as citrate reduced Ag colloid, which have been used for single-molecule SERS.[55] To clearly demonstrate the advantages of the SHAANs, the SERS of MBA on SHAANs were compared with solid nanostars synthesised following a well-recognized surfactant-free approach from literature and with surfactant free hollow Ag-Au nanocubes produced via an in-house procedure derived from literature.[54,57] As shown in Figure S10, the hollow nanocubes were ca. 250 nm in diameter, which was similar to the diameter of the cubic hollow core of the SHAANs, while the solid nanostars were ca. 50 nm in diameter. As shown in Figure S11, the SERS signals of MBA were clearly the strongest on SHAANs. More specifically, under exactly the same unaggregated experimental conditions, no signals of MBA were observed from the hollow nanocubes, while the SERS signals of MBA obtained from the conventional solid nanostars were >2× weaker than the signals of MBA obtained from SHAANs.

It is important to note that the strong AEF of SHAANs arises not only from their excellent plasmonic properties but also from their chemically exposed surfaces. More specifically, the surfactant-free synthesis means that the only chemical species adsorbed on the surface of the SHAANs were Cl$^-$ ions originating from HAuCl$_4$ and NaCl. It is known that charged ligands, such as Cl$^-$, only occupy ca. 20% of the surface of Ag/Au nanoparticles due to intermolecular electrostatic repulsion.[58] As a result, the blank SERS signals of the pristine SHAANs colloids only showed a single weak Au-Cl vibration band (Figure S12). This means the surface of SHAANs are much more accessible to analyte molecules than conventional spiky nanostars which are covered in strongly adsorbed ligands, such as PVP or poly(ethylene glycol) methyl ether thiol. The lack of capping agents in surfactant-free syntheses does lead to issues in the long-term stability of the product nanoparticle colloids, since there are no capping ligands to protect the nanoparticles from aggregation. Indeed, it was found that the pristine SHAANs colloids were very unstable against aggregation and typically precipitated as aggregates from solution within 1 h. The challenge is to be able to stabilize the SHAANs colloid without compromising the surface-accessibility of the particles. This was not a problem since another advantage of the surfactant-free environment is that it allowed us to use weakly adsorbed polymeric molecules, such as HEC polymer, as stabilizing agents.[59-61] This allowed the SHAANs to remain highly stable for >1 month as characterized by UV-vis spectroscopy (Figure S13) while retaining their surface-accessibility for further applications. Interestingly, SERS monitoring of the same batch of SHAANs revealed that their SERS activity decreased slightly, by ca. 2×, during the first day of storage (Figure S14). This is most likely due to oxidation of the thin Ag surface-layer rather than aggregation or Ostwald ripening of the SHAANs since the extinction spectrum of the colloid remained unchanged.[17,62] More importantly, the long-term SERS kinetics showed that the slight decrease in SERS activity was only observed during the first day of storage, after which the SHAANs remained highly stable plasmonically for >1 month. The high chemical stability of SHAANs against further oxidation is likely due to SHAANs being mainly composed of Ag-Au alloy, which is consistent with previous plasmonic studies comparing Ag and Ag-Au nanocubes.[62]

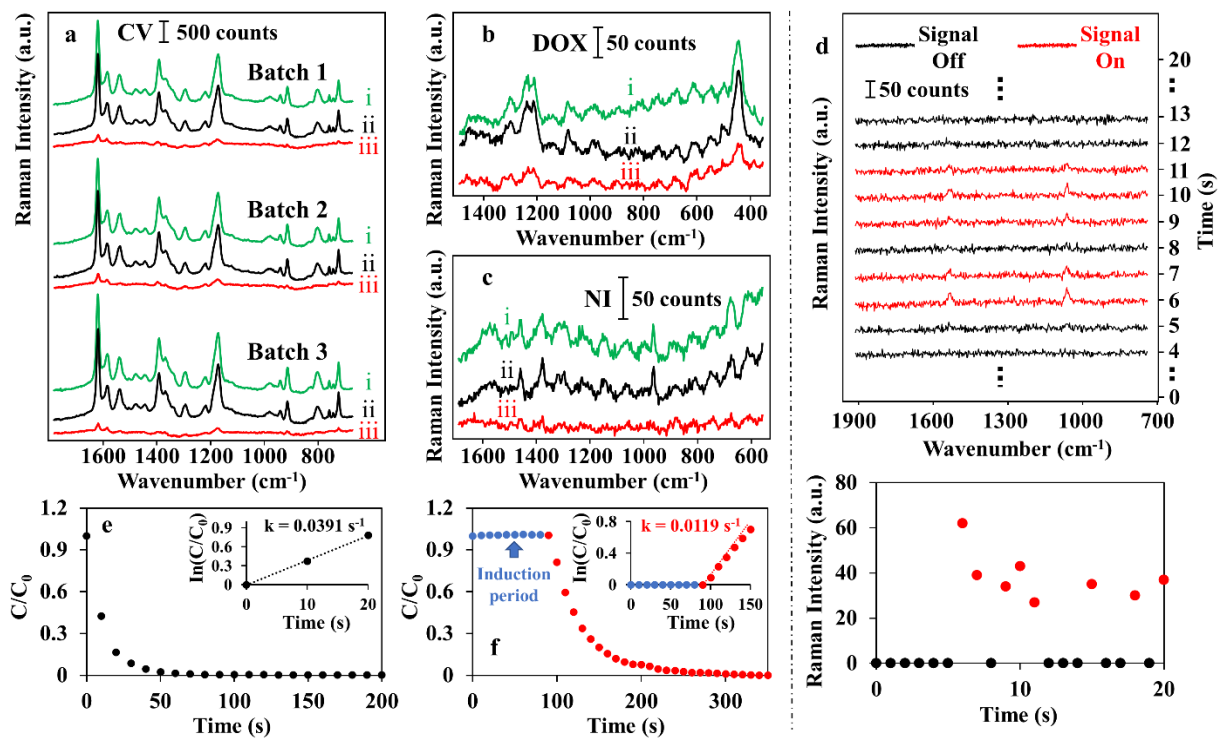

**Figure 5**. (a) SERS spectra of $10^{-6}$ M crystal violet (CV), (b) $10^{-5}$ M Doxorubicin (DOX) and (c) $10^{-5}$ M Niraparib (NI) recorded on pristine SHAANs colloid (i), hydroxyethyl cellulose (HEC) stabilized SHAANs colloid (ii) and PVP stabilized SHAANs colloid (iii). (d) The time-dependent SERS spectra showing single SHAANs moving into and out of the probed laser volume. Spectra with detectable SERS signals of the MBA analyte at 1045 cm$^{-1}$ are highlighted in red. The plot below shows the change in SERS signal for an extended period of 20 seconds with red dots representing the appearance of MBA signal and black dots representing the disappearance of MBA signal. (e-f) Plots showing the decrease in concentration of 4-nitrophenol with time in the NaBH$_4$ reduction of 4-nitrophenol catalysed by HEC (e) and PVP-stabilized SHAANs (f). Insets show the rate of the reactions.

Figure 5a compares the SERS activity of three batches of pristine, HEC-stabilized and PVP-stabilized SHAANs colloids for the detection of a model SERS analyte, crystal violet (CV). As shown in spectra sets i-ii, the SERS signals of CV were similarly intense on HEC-SHAANs and pristine SHAANs. In contrast, since the Cl$^-$, which must adsorb onto the enhancing surface to facilitate the adsorption of the cationic CV dye, cannot displace PVP, a significant drop in SERS intensity by ca. 10× was observed when PVP-SHAANs (Figure 5a, spectrum iii) were used as the enhancing substrate. It is worth noting that CV is not a special case. The exposed surface of SHAANs also provides significant advantages in SERS detection of important real-life targets. For example, Figure 5b-c compares the SERS signals of two anticancer drugs, Niraparib and Doxorubicin, obtained using pristine, HEC-SHAANs and PVP-SHAANs at a final drug concentration of $10^{-5}$ M. For both anticancer drugs, the SERS signals were intense on pristine/HEC-SHAANs but barely visible on PVP-SHAANs, which emphasizes the importance of surface accessibility in SERS monitoring of real-life analytes.

The outstanding AEF of SHAANs also suggests that they might be suitable for single particle SERS measurements, which is an important property in both fundamental and applied studies, such as SERS sensing/imaging in biological samples.[63-64] To test this, HEC-SHAANs colloid was diluted, labelled with MBA, which is commonly used as a Raman tag molecule for biological studies, and Raman probed using a confocal Raman microscope with a diffraction-limited probe diameter of 1.04 µm. Figure 5d shows a subset of the SERS spectra from a total of 20 spectra series recorded continuously at 1 s intervals. Even though MBA is a non-resonant analyte and the accumulation time was short, the signal of MBA was still intense and was measured to be 38±11 counts using the characteristic peak at 1045 cm$^{-1}$. More importantly, the MBA signals showed an on/off pattern, which is associated with the movement of the enhancing particle-substrate moving into and out of the probed volume due to Brownian motion. Since within this diluted SHAANs colloid there is only a single particle every 2000 µm$^3$ (see supporting info for details), which was more than 2000× larger than the probe volume (0.7 µm$^3$) of the confocal microscope, this meant that the consistently observed on/off signals of MBA most likely arose from a single SHAAN in the probe laser acting as a single-particle SERS probe. To validate that the SHAANs were active as single particles the effect of colloidal aggregation on the plasmonic properties of SHAANs were studied by performing contrast SERS kinetic experiments using pristine and HEC-SHAANs. As shown in Figure S15, the SERS signals obtained from the pristine SHAANs decreased almost linearly as they aggregated naturally without any visible precipitation while the SERS signals of the unaggregated HEC-SHAANs remained constant over the same period of time. This shows that aggregation has a negative impact on the SERS activity of the SHAANs which is consistent with previous reports regarding star-type Au particles,[65-66] and more importantly, argues against attributing the intermittent signals in Figure 5d to aggregates. The SERS properties of the SHAANs functionalized with MBA Raman tags were also studied in artificial serum, which contained 3.3 wt.% albumin, to mimic a real-life environment. As shown in Figure S16, the strong SERS signals of MBA from the SHAANs were fully retained, which demonstrates their potential in biochemical applications, for example as pH reporters.

The HEC-SHAANs also showed superior catalytic properties to PVP-SHAANs. As shown in Figure 5e, the reduction of 4-nitrophenol by NaBH$_4$ occurred immediately when catalysed by HEC-stabilized SHAANs, as measured by UV-vis spectroscopy (Figure S17). In contrast, for PVP-stabilized SHAANs there was a delay of ca. 90 s before any reaction could be detected (Figure 5f and Figure S18). In addition, the average rate constant for the reaction catalysed by the HEC-SHAANs ($k$=0.0406±0.0015 s$^{-1}$) was >3.8× faster compared to their PVP-stabilized counterparts ($k$=0.0105±0.0015 s$^{-1}$), as shown in the insets of Figures 5e-f. This can be attributed to the need for surface reconstruction of PVP-SHAANs when acting as catalysts due to the strongly adsorbed PVP capping on the particles' surfaces.[67]

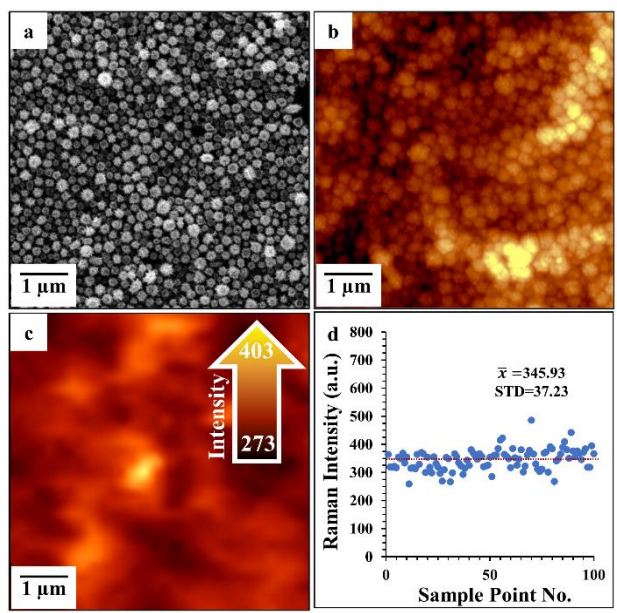

**Figure 6**. (a) A SEM image of a SHAANs SENS showing a monolayer of densely packed SHAANs on the surface. (b) Atomic force microscopy (AFM) and (c) Raman mapping images across a 5 µm × 5 µm area on the SHAANs SENS labelled with MBA. (d) The Raman intensity of the MBA peak at 1045 cm$^{-1}$ obtained from 100 random spots on the Raman mapping image.

The high yield and excellent properties of the SHAANs also makes them promising candidates for the construction of multidimensional nanomaterials with advanced functionalities. For illustration, we show that HEC-SHAANs can be readily used as the building-block for the construction of surface-exposed nanoparticle sheets (SENSs) via an interfacial self-assembly technique previously reported by our group.[68] Figure 6 shows typical SEM and SERS results obtained from SHAANs SENSs. As shown in Figure 6a-b, the SHAANs in SENS are held as a densely packed 2-dimensional monolayer anchored on the surface of the polystyrene. Importantly, the surface of the HEC-SHAANs in SENSs are both chemically and physically exposed for interactions in applications. Figure 6c shows a 5 µm × 5 µm Raman map of the HEC-SHAANs SENS with MBA adsorbed on the particle surface. As shown in Figure 6d, the assembled SHAANs remained highly plasmonically active and the relative standard deviation of the SERS intensity of the MBA peak at 1045 cm$^{-1}$ across 100 randomly selected points within this mapped area was very low at just 10.8%, which demonstrates the potential of HEC-SHAANs as a plasmonic building block for constructing uniform and active plasmonic devices.

**Conclusions**

In conclusion, we have demonstrated a rapid, mild and surfactant-free one-pot synthesis to prepare spiky Ag-Au nanoparticles (SHAANs) with near 100% morphological yield. The as-synthesized nanoparticles have a hollow interior with a dense covering of narrow spikes irradiating from the surface which gave rise to distinct catalytic and plasmonic properties. Moreover, the fact that the synthesis was surfactant-free meant that the surfaces of the SHAANs were chemically exposed, which allowed the use of HEC as a weakly bound stabilizing agent to ensure colloidal stability while retaining surface accessibility for further applications. As a result, the HEC-SHAANs were not only highly stable but also exhibited greatly enhanced functionalities as both SERS substrates and catalysts when compared against their surfactant-capped counterparts. The combination of easy preparation and outstanding functionality means that SHAANs have potential to be exploited in a broad range of applications as both functional materials in their own right and as building-block for functional materials with advanced properties.

**AUTHOR INFORMATION**

**Corresponding Author**

Yikai Xu

E-mail: yxu18@qub.ac.uk

Steven E. J. Bell

E-mail: s.bell@qub.ac.uk

**Author Contributions**

The manuscript was written through contributions of all authors. † These authors contributed equally.

**Funding Sources**

ZY, CL, YX and SB acknowledges the University Special Research Scholarship (Q.U.B) for support. YX acknowledges The Leverhulme Trust Early Career Fellowship (grant ECF2020703) and Royal Society of Chemistry Researcher Mobility Grant for financial support. MC and SB acknowledges the EPSRC grant EP/P034063/1 for support. CH and AG acknowledge support from the UK Catalysis Hub, funded by EPSRC grant EP/R027129/1. SJH and ML acknowledge funding from the EPSRC (UK) (grants EP/M010619/1, EP/S021531/1, EP/P009050/1) and the European Commission H2020 ERC Starter grant EvoluTEM (715502). This work was supported by the Henry Royce Institute for Advanced Materials, funded through EPSRC grants EP/R00661X/1, EP/S019367/1, EP/P025021/1 and EP/P025498/1.

**ACKNOWLEDGMENT**

The authors would also like to thank: Dr. Conor Byrne and Dr. Alex Walton from The University of Manchester for measuring the XPS samples; Prof. Zhong-Qun Tian, Prof. Jian-Feng Li, Dr. Yue-Jiao Zhang from Xiamen University for useful discussions; Ms. Harmke S. Siebe from University of Groningen for assisting with HEC-SHAANs preparations.

**ABBREVIATIONS**

PVP, polyvinylpyrrolidone; CTAB, cetyltrimethylammonium bromide; SHAANs, spiky hollow Au-Ag nanoparticles;

**Supporting Information**

**Materials and methods**

**Materials:** Silver nitrate, tetrachloroauric (III) acid trihydrate, sodium chloride, L-ascorbic acid, polyvinyl pyrrolidone (50,000 M.W.), hydroxyethyl cellulose (250,000 M.W.), 4-mercaptobenzoic acid, crystal violet, dichloromethane, ethanol, tetrabutylammonium nitrate and polystyrene (192,000 M.W.) were purchased from Sigma Aldrich and used as received. Doxorubicin and Niraparib were purchased from Selleckchem.com Ltd. All experiments used low TOC (<3.0 ppb) 18.2 MΩcm$^{-1}$ water.

**Synthesis of SHAANs, solid nanostars and hollow nanocubes**: In a typical synthesis of SHAANs, 60 µL of 1 M NaCl solution and 0.5 mL of 3 mM AgNO$_3$ solution was added in this sequence to 50 mL of 0.25 mM HAuCl$_4$ aqueous solution under vigorous magnetic stirring. This mixture was incubated at room temperature for 10 min under stirring. After incubation, 0.25 mL of 100 mM L-ascorbic acid solution was quickly injected into the stirred solution. After injection of L-ascorbic acid, the mixture was left to react for 15 seconds, during which the color of reaction mixture changed from light yellow to colorless, and finally to light blue, which indicated the formation of SHAANs. After the reaction, 5 mL of 1 $_{wt.}$% HEC (or 5 mL of 0.1 $_{wt.}$% PVP) solution was added to the SHAANs colloid and stirred for 5 mins to form HEC/PVP-SHAANs colloids, which remained stable at room-temperature for at least 1 month. The concentration of a typical batch of SHAANs was ca. 10$^9$ particles/mL as measured with dynamic light scattering coupled with nanoparticle tracking analysis using a Nanosight NS300.

Solid nanostars were synthesized following a protocol in literature.[57] Immediately after synthesis, 1 mL of 1 wt.% HEC (or 1 mL of 0.1 wt.% PVP) was introduced to the colloid and stirred for 5 mins to form stable HEC/PVP-nanostar colloids, which remained stable for at least 1 month.

Hollow nanocubes were synthesized using a protocol derived from literature.[54] In a typical synthesis of hollow nanocubes, 45 μL of 1 wt.% $HauCl_4·3H_2O$ stock solution was mixed with 10 mL of DDI water under vigorous magnetic stirring for 1 min. After this, 170 μL of 6 mM $AgNO_3$ solution was injected into the stirred solution. After 1 min of incubation, 500 μL of 0.1 M AA solution was injected into the stirred solution, which led to the formation of a light purple colloid within 10 s. The colloid was allowed react for 30s. After this, 1 mL of 1 wt.% HEC (or 1 mL of 0.1 wt.% PVP) was introduced to the colloid and stirred for 5 mins, to form stable colloids, which remained stable for at least 1 month. To purified colloid and remove spherical by-products, the stabilized colloid was centrifuged at 1000 rcf for 15 mins. All but 500 μL of the pink supernatant was removed, and the colloid was redispersed in DDI water.

**Preparation of SHAANs SENSs**: The detailed process can be found in our previous publication.[68] Briefly, 20 mL of PVP/HEC-stabilized SHAANs colloid was first concentrated by a factor of 10 using centrifugation. Then a mixture of 2 mL of the concentrated PVP/HEC-SHAANs, 0.1 mL of 1 mM tetrabutylammonium nitrate, and 3 mL of dichloromethane containing 0.18 g of dissolve polystyrene was vigorously shaken by hand for 1 minute to create water-dichloromethane emulsion droplets onto which SHAANs self-assembled. After shaking, the mixture was poured immediately into a polypropylene petri dish which was quickly covered with a lid to allow the gradual coalescence of emulsion droplets. The complete coalescence of emulsion droplets led to the formation of a SHAANs 2D array at the water-dichloromethane interface. After this the lid was removed to allow the evaporation of dichloromethane, which led to the precipitation of dissolved polystyrene and the formation of a SHAANs SENS at the water-air interface.

**Microscopy characterizations**: SEM was performed with a Quanta FEG 250 at an acceleration voltage of 20 keV under high chamber vacuum ($8×10^{-5}$ mbar) with standard copper tape as background. STEM-HAADF and EDX spectrum imaging was performed using an aberration corrected FEI Titan G2 80-200 ChemiSTEM. The microscope was operated at 200 keV with a beam current of 180 pA. STEM-HAADF imaging was acquired using the Tecnai Imaging and Analysis (TIA) software and EDX data acquired using an FEI Super-X quad silicon drift detector, which has a collection angle of ~0.7 sr, processed using Bruker Esprit software and Python-based modules. AFM AC mode was used for mapping an area of 5 μm × 5 μm on a SHAANs SENS using a WItec Alpha 300 Raman microscope with a driving amplitude of 0.12 Vpp and a setpoint of 4.5 V. 300 points were taken per line and 300 lines were taken per image.

**XPS spectroscopy:** XPS was performed using a SPECS UHV system (base pressure $5×10^{-10}$ mbar), with a monochromated Al Kα radiation X-ray source (1486.6 eV) operating at 120 W, and analysed using a PHOIBOS 150 NAP electron energy analyser. Samples were drop cast on a silicon substrate at 90 °C. Survey scans were acquired at an analyser pass energy of 60 eV and high-resolution narrow scans were performed at a constant pass energy of 20 eV in 0.1 eV steps. An electron flood gun was used during the analysis (45 μA at 5 eV). The spectra were analysed using CasaXPS and corrected for charging using the C 1s feature at 284.8 eV.

**Raman spectroscopy:** Liquid-state SERS detection were performed by adding 20 μL of analyte solution ($10^{-5}$ M CV, MBA or $10^{-4}$ M DOX, NI) to 200 μL of (pristine or PVP/HEC stabilized) nanoparticle colloid. All the spectra lines shown in the Figures were averaged from 3 independent accumulations.

SERS studies of CV and SERS studies comparing the plasmonic activities of unaggregated solid nanostars, hollow nanocubes and SHAANs were performed on a Avalon R2 Ramanstation equipped with a 785 nm laser with 30 mW laser power. The total accumulation time for each SERS spectrum was 30 seconds. To fairly assess the plasmonic activities of unaggregated solid nanostars, hollow nanocubes and SHAANs, the nanostars, hollow-nanocubes and SHAANs were all stabilized using PVP. The concentrations of the particle colloids were adjusted to $10^9$ particles/mL based on concentration measurements performed with a Nanosight particle tracking system. To mitigate the effect of the PVP capping to SERS analysis, 4-MBA was used, and the SERS of the three types of colloids containing $10^{-6}$ M of 4-MBA were measured.

SERS of DOX and NI and AEF estimations of the SHAANs using MBA were performed using a WItec Alpha 300 Raman microscope equipped with a 10× lens and a 785 nm laser with 30 mW laser power. The total accumulation time for each SERS spectrum was 30 seconds. Single-particle SERS samples were prepared by first adding 20 μL of $10^{-4}$ M MBA to 200 μL of HEC-stabilized SHAANs. 200 μL of water was then added to dilute the MBA labelled SHAANs to half concentration. Time-dependent SERS measurements were performed on a WItec Alpha 300 Raman microscope equipped with a 100× lens (with a spot size of 1 μm) and a 785 nm laser with 30 mW laser power. The accumulation time for each of the 20 continuous measurements was 1 second.

Raman mapping samples were prepared by first dipping a piece of SHAANs SENS into $10^{-5}$ M MBA solution for 1 minute to label the surface with MBA. The SENSs were then washed with ethanol to remove excessive MBA and then dried at room temperature. Raman mapping of SHAANs SENSs was performed on a 5 μm × 5 μm area using a WItec Alpha 300 Raman microscope equipped with a 785 nm laser and a 100× lens (with a resolution of 500 nm). The laser power was 0.28 mW and the accumulation time for each measurement was 10 seconds. 28 points were taken per line with a 180 nm interval and 28 lines were taken per image.

SERS of SHAANs in artificial serum were performed using a Perkin Elmer RamanMicro 200F microscope equiped with a 785 nm laser with 40 mW laser power and 60 μm spotsize. The accumulation time was 50 s.

**Catalytic reduction of 4-nitrophenol by NaBH$_4$**: 1 mL of 0.1 mM 4-nitrophenol was first mixed with 0.5 mL of 50 mM NaBH$_4$ in a quartz cuvette. To initiate the reduction reaction, 1 mL of SHAANs colloid was added to the solution mixture. The reaction was monitored by UV-vis spectroscopy on an Agilent 8453 photodiode array UV–vis spectrophotometer. UV-vis spectra were taken with 7-second and 10-second intervals for reactions catalysed by HEC and PVP stabilized SHAANs, respectively. The intensity of the band at 400 nm corresponding to 4-nitrophenol was monitored.

**Computational method:** FEM implemented in COMSOL Multyphysics 5.5 was used to model the electrodynamics of different Ag-Au alloy (1:9) structures. Geometrical parameters used in these simulations were chosen on the basis of experimental data. The edge length of the cube was set to 140 nm. The shell thickness was set to 20 nm. 25 spikes were placed on each cube face and represented by a rounded tip cone with length variable between 20 to 40 nm. The radii of the top and bottom surfaces of the spike were set to be 5 and 10 nm, respectively and spikes separated by a 5 nm gap. The Wave Optics module was used in all simulations. A spherical volume of water (1600 nm radius) around the structure was also included in the model. The external layer of this volume (thickness 1000 nm) was set as perfectly matched layer domain to act as an absorber of the scattered field. A volume of water was also added in the cavity of the hollow structures as well. Physically controlled meshing was used with ultrafine settings for lossy materials. Dielectric function data ranging from 400 nm to 1200 nm for a Ag-Au alloy (1:9) was taken from Rioux et al.[69] A scattered-field plane wave source with wavelength ranging from 400 to 1200 nm was selected to estimate the interaction between propagating plane wave and metallic nanostructures. Each structure was excited by a linearly polarized electromagnetic field, polarization axes perpendicular and parallel to the cube and spike axis were used.

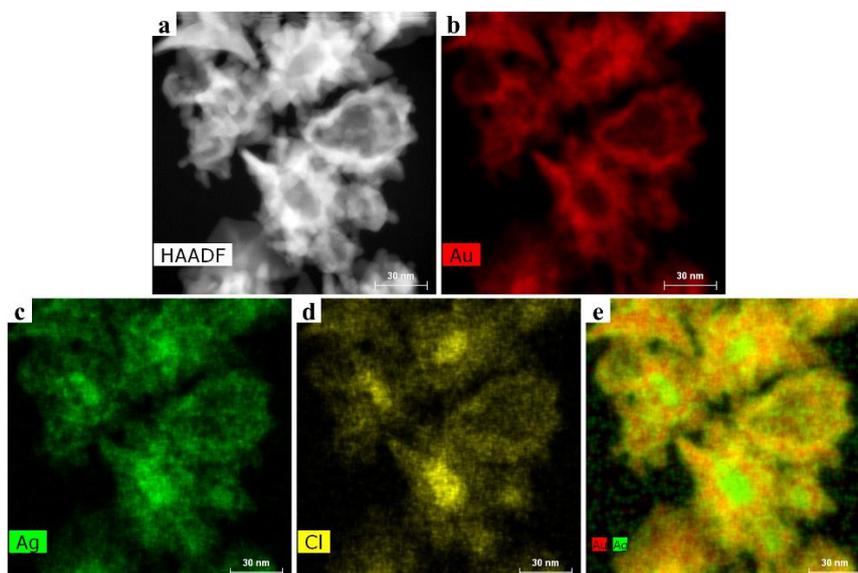

**Figure S1.** (a) Scanning transmission electronic microscopy high-angle annular dark field (STEM-HAADF) imaging of nanostars produced with 0 min of incubation. (b-d) Energy dispersive X-ray (EDX) mapping of the nanostars which reveals AgCl cores.

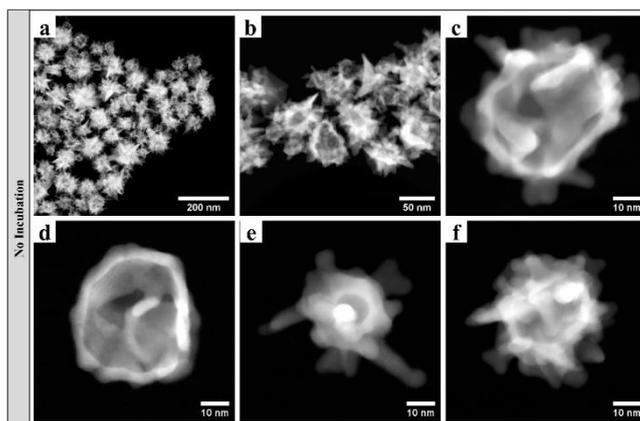

**Figure S2.** STEM-HAADF analysis for a non-incubated specimen showing a range of particle morphologies, some with a smooth exterior surface and others with spikes.

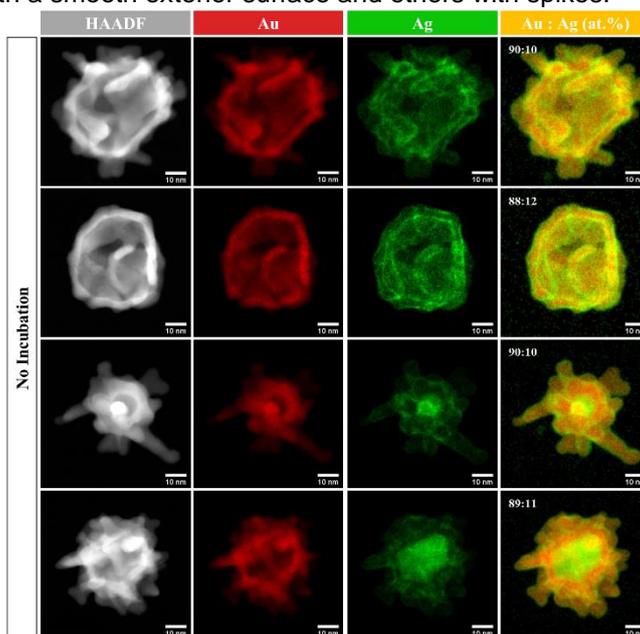

**Figure S3.** STEM-HAADF and EDX characterization of the different nanostar species within the non-incubated sample showing that the atomic ratios of Au:Ag for individual nanostars (ratios inset to composite maps in the right-hand column) were ca. 9:1.

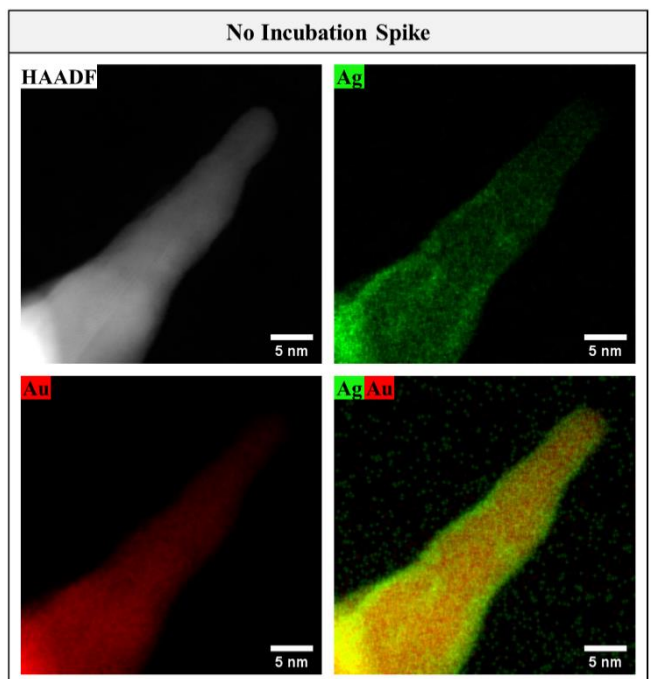

**Figure S4.** STEM-HAADF and EDX characterization of the tip of a nanostar produced with 0 mins of incubation showing a few-nanometer thick Ag-rich surface layer.

In general, a series of chemical reactions is triggered simultaneously when L-Ascorbic Acid is added into a mixture of AgNO$_3$ and HAuCl$_4$, as shown in **Table S1** below:

$$[AuCl_4]^- \leftrightarrows [AuCl_3OH]^- \leftrightarrows [AuCl_2(OH)_2]^- \leftrightarrows [AuCl(OH)_3]^- \leftrightarrows [Au(OH)_4]^- \qquad 1$$

$$[AuCl_{4-x}OH_x]^- + AA \xrightarrow{Ag^+} dehydroAA + Au \qquad 2$$
$$Ag + [AuCl_{4-x}OH_x]^- \rightarrow Au + Ag^+ \qquad 3$$
$$Ag^+(free\ silver\ and\ silver\ chloride) + AA \rightarrow dehydroAA + Ag \qquad 4$$
$$Ag^+ + Cl^- \leftrightarrows AgCl \qquad 5$$

The correct incubation time controls the size and morphology of the AgCl seeds as well as the concentration of each reactant, which in turn provides the correct balance between reactions 1-5 to generate SHAANs.

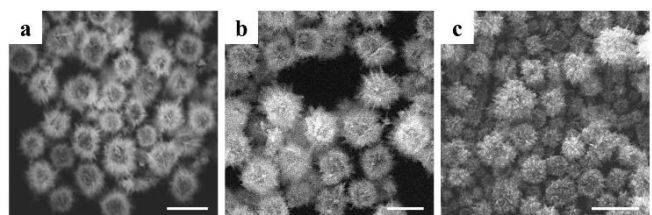

**Figure S5.** Scanning electron microscopy (SEM) images of three different batches of SHAANs. All scale bars correspond to 250 nm.

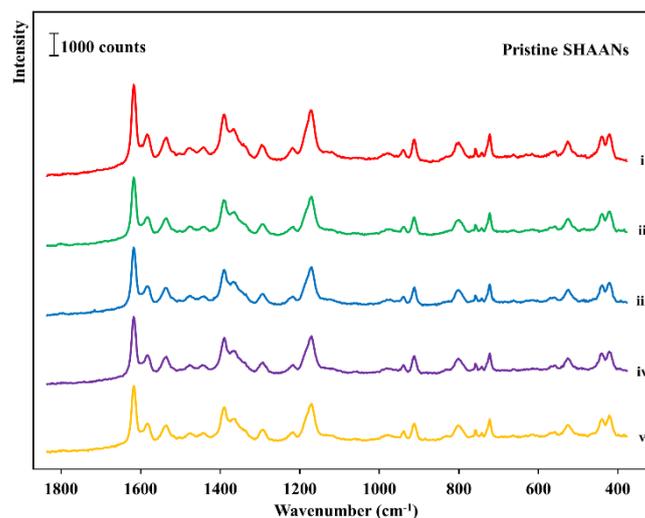

**Figure S6.** SERS spectra of 10$^{-6}$ M crystal violet obtained using different batches of pristine SHAANs.

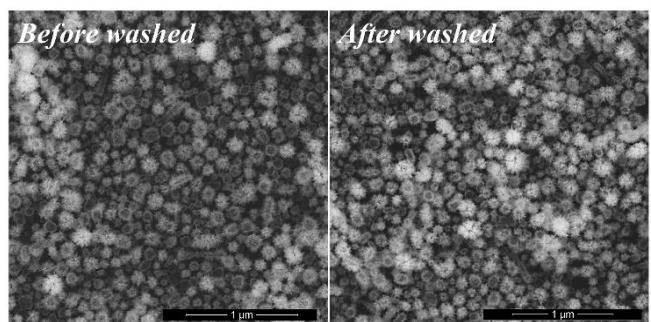

**Figure S7.** SEM images of SHAANs before and after washing with NH$_3$·H$_2$O. The scale bars correspond to 1 μm.

**Table S2.** Elemental composition from XPS in at.% for the 200 nm SHAANS sample. Na, Cl, C, O and N can be attributed to NaCl, NO$_3^-$ and HEC introduced during the synthesis of SHAANs and are part of the dried sample analysed by XPS.

| Element | Composition (at.%) |
|---|---|
| Au 4f | 0.33 |
| Ag 3d | 0.20 |
| Cl 2p | 2.95 |
| Na 1s | 2.13 |
| C 1s | 54.65 |
| O 1s | 26.51 |
| N 1s | 3.94 |

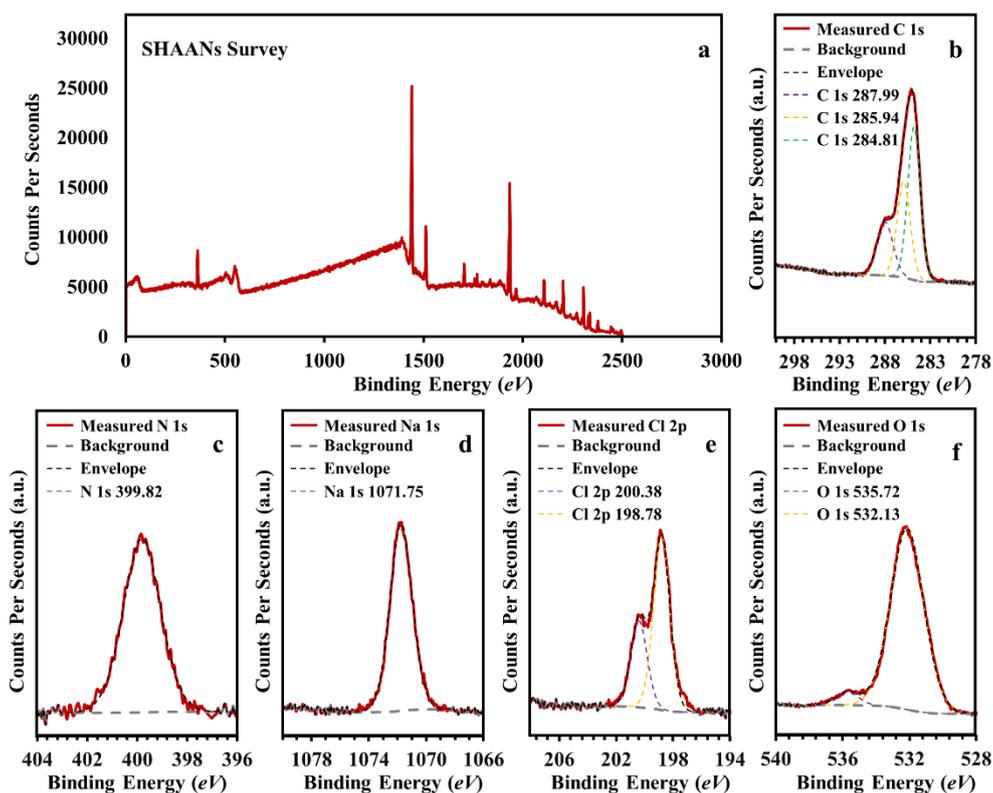

**Figure S8.** XPS spectra of the 200nm SHAANS sample: (a) survey, (b) C 1s, (c) N 1s, (d) Na 1s, (e) Cl 2p, (f) O 1s.

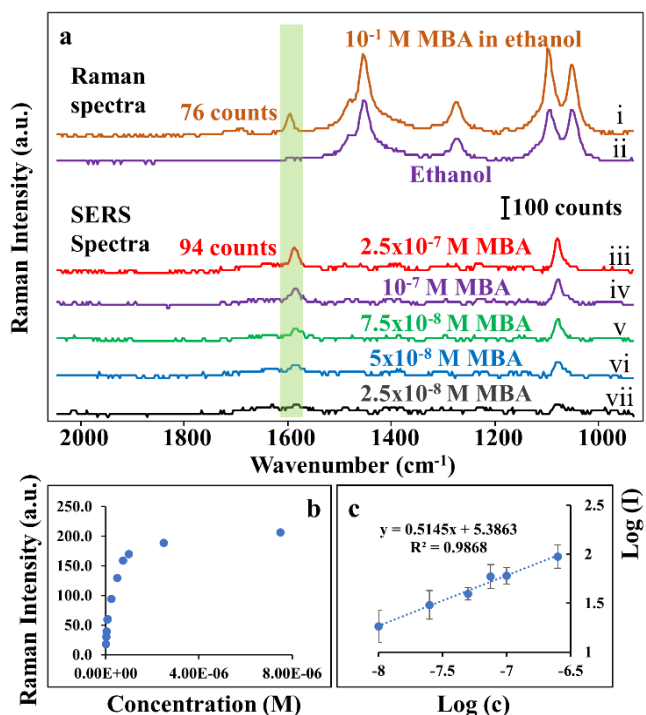

**Figure S9.** (a) Raman spectra of MBA ($10^{-1}$ M) dissolved in ethanol (i) and pure ethanol (ii). SERS spectra of MBA at different concentrations obtained using HEC stabilized SHAANs as SERS substrates (iii-vii). (b) Plot showing the change in MBA SERS intensity against its concentration. (c) Plot showing the linear calibration curve at concentrations between $10^{-8}$ M to $2.5 \times 10^{-8}$ M.

The analytical enhancement factor (AEF) of a NP colloid can be estimated through the ratio between the SERS ($I_{SERS}$) and Raman ($I_{Raman}$) intensity for the selected mode of a given analyte under identical experimental conditions (instrument, accumulation time, etc.) following the equation below.

$$AEF = {I_{SERS}C_{Raman}}/{I_{Raman}C_{SERS}} \qquad \text{Equation 1}$$

where $C_{SERS}$ and $C_{Raman}$ correspond to the concentrations of the analyte in SERS and Raman measurements, respectively. Therefore, by using the intensity of the band at 1590 cm$^{-1}$ as highlighted in Figure 6.4, the AEF value for SHAANs is

$$AEF = {94 \times 10^{-1} M}/{76 \times 2.5 \times 10^{-7} M} \approx 4.9 \times 10^5 \qquad \text{Equation 2}$$

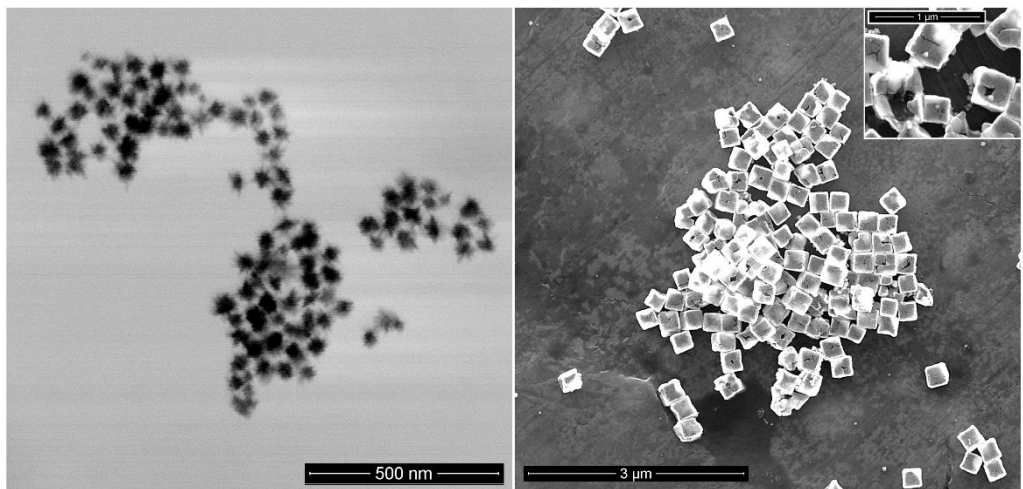

**Figure S10.** SEM image of solid nanostars ca. 50 nm in diameter and hollow nanocubes ca. 250 nm in diameter. The inset shows the hollow nanocubes at higher magnification showing the hollow cavity.

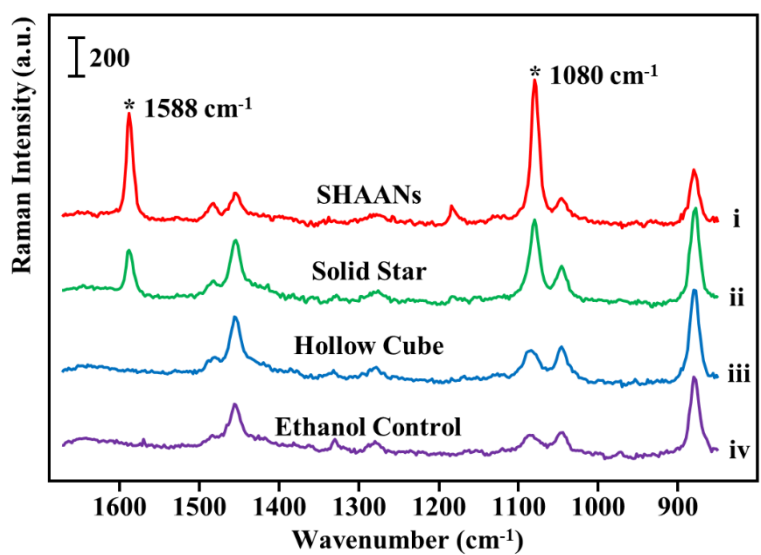

**Figure S11.** Spectra set comparing the SERS signals of 4-MBA obtained from unaggregated SHAANs, solid stars and hollow nanocubes under identical experimental conditions (see experimental section).

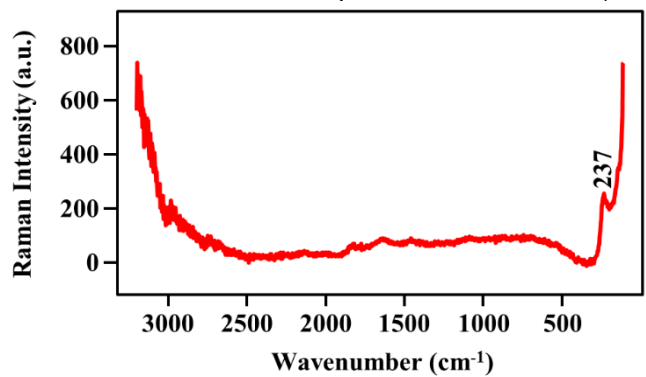

**Figure S12.** SERS background of pristine SHAANs showing a single peak corresponding to Au-Cl vibration.

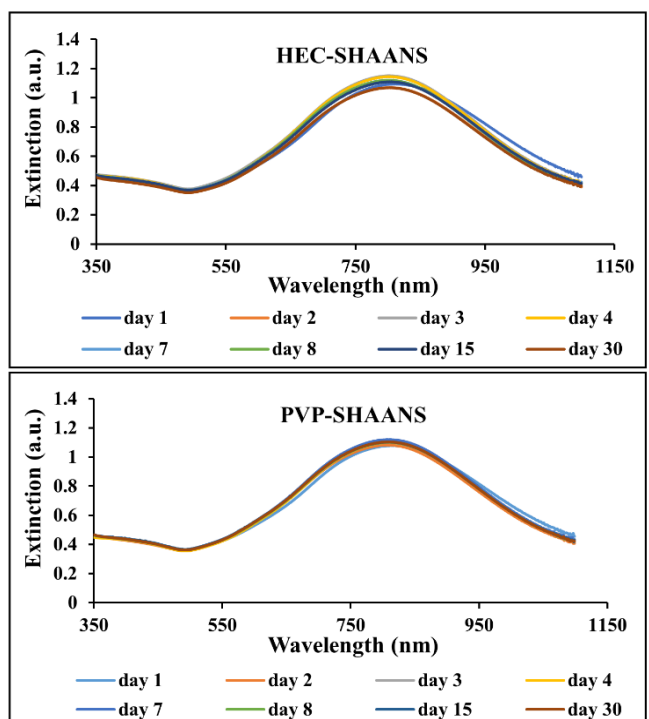

**Figure S13.** Extinction spectra of SHAANs produced from the same batch of synthesis stabilized by HEC and PVP collected over the course of 30 days. Plots for pristine SHAANs are not shown since the particles aggregates and quickly.

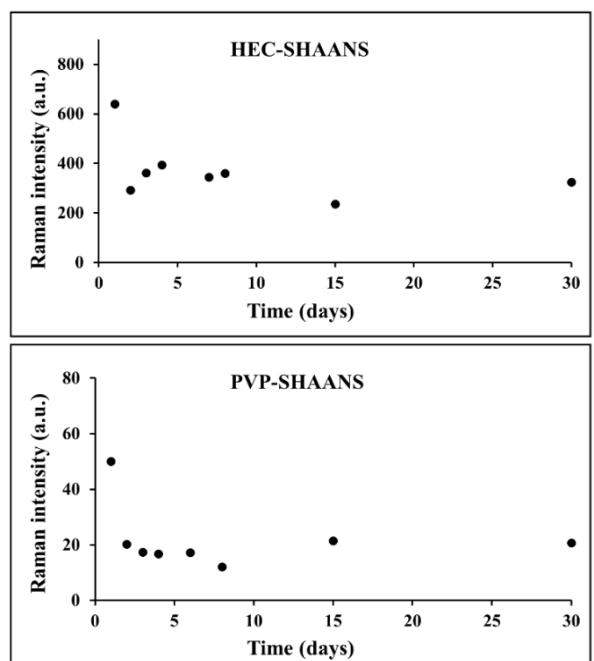

**Figure 14.** Plot showing the SERS intensity of crystal violet Raman tag measured using HEC-/PVP-SHAANs colloids as the enhancing material over the course of 30 days.

**Calculation for single particle SERS**

The diameter of the laser beam can be calculated by the following equation:

$$d = \frac{1.22 \lambda_{laser}}{NA} = \frac{1.22 \times 785}{0.9} \approx 1 \; \mu m \quad \quad \text{Equation 3}$$

The confocal volume $V_c$ can be calculated by the following equation:

$$V_c = \pi^{1.5} k \left(\frac{d}{2}\right)^3 = \pi^{1.5} \times 2.34 \times 0.5^3 \approx 0.695 \ \mu m^3 \qquad \text{Equation 4}$$

where $k$ is the resolution of the objective in the z direction divided by the resolution in the x, y plane, and $d$ is the diameter of the laser beam.

The concentration of the diluted SHAANs is ca. $5 \times 10^8$ particle/mL, which translates to 1 particle per 2000 $\mu m^3$, whereas the confocal volume is only $0.695 \ \mu m^3$, it is most likely that there is only one SHAANs in the probe volume during the SERS measurement.

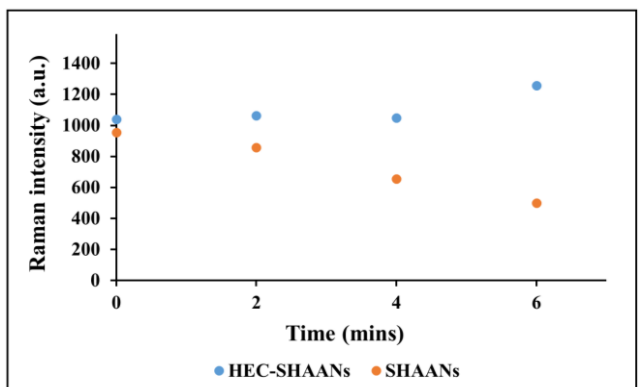

**Figure S15.** Plot showing the SERS intensity of thiophenol Raman tag measured using HEC-SHAANs and pristine SHAANs colloids as the enhancing material over 6 mins.

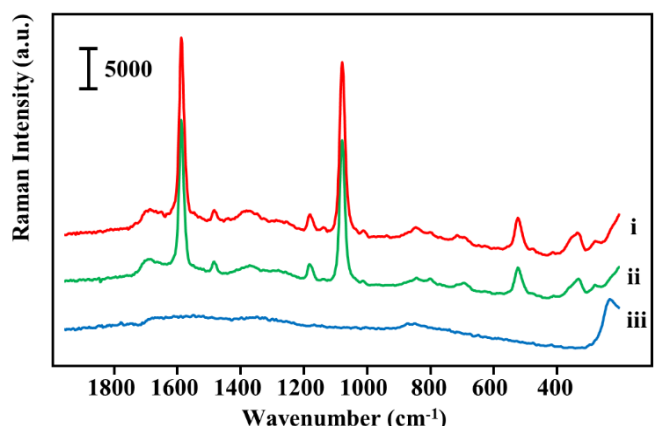

**Figure S16.** SERS spectra set showing blank spectrum of HEC-SHAANs (iii), SERS spectrum of HEC-SHAANs with $10^{-5}$ M of MBA (ii), SERS spectrum of HEC-SHAANs with $10^{-5}$ M MBA and 3.3 wt.% albumin solution. The spectra were acquired from the same batch of HEC-SHAANs.

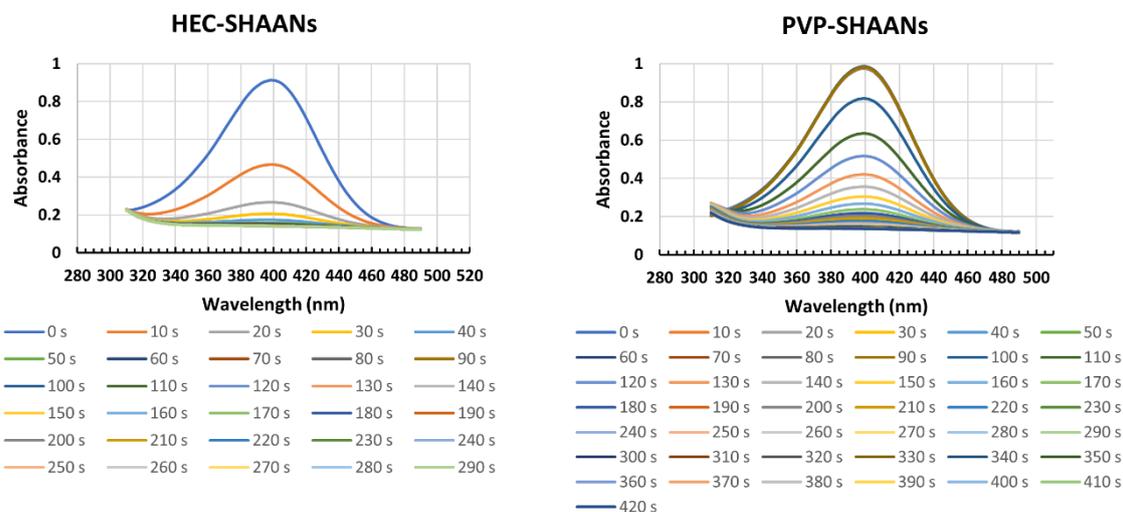

**Figure S17.** UV-vis spectra sets showing the decrease in concentration of 4-nitrophenol with time in the $NaBH_4$ reduction of 4-nitrophenol catalysed by HEC and PVP-stabilized SHAANs.

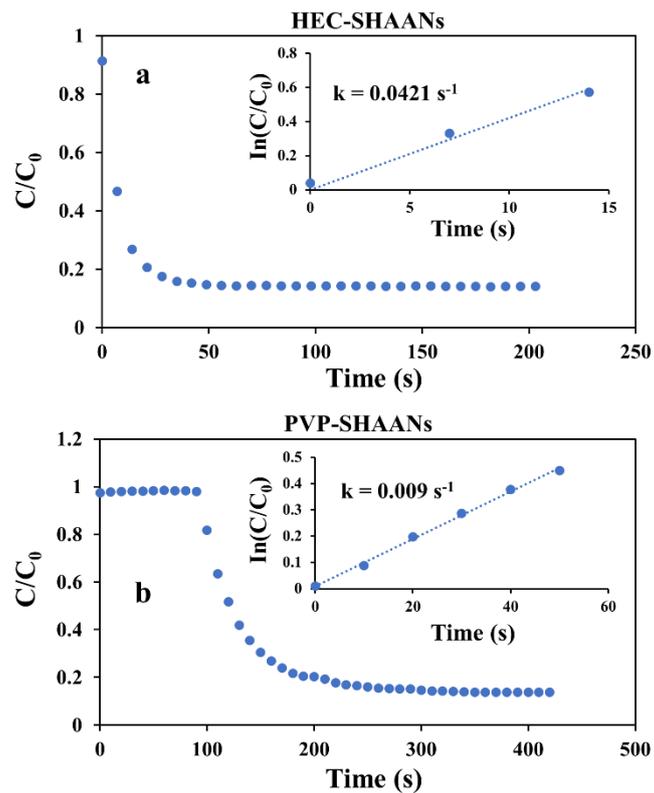

**Figure S18.** Plots showing the decrease in concentration of 4-nitrophenol with time in the NaBH$_4$ reduction of 4-nitrophenol catalysed by a different batch of (a) HEC and (b) PVP-stabilized SHAANs. Insets show the rate of the reactions.